\documentclass[aps,english,prc,preprintnumbers,fleqn,showkeys,
superscriptaddress,floatfix]{revtex4}

\usepackage[T1]{fontenc}
\usepackage[latin9]{inputenc}
\usepackage{wrapfig}
\usepackage{calc}
\usepackage{amsmath}
\usepackage{graphicx}
\usepackage{float}
\usepackage{calc}
\usepackage{amstext}

\def\beq{\begin{equation}}
\def\eeq{\end{equation}}
\def\bea{\begin{eqnarray}}
\def\eea{\end{eqnarray}}
\def\beqa{\begin{equation}\begin{array}{l}}
\def\eeqa{\end{array}\end{equation}}
\def\eqlab#1{\label{eq:#1}}
\def\figlab#1{\label{fig:#1}}

\def\Eqref#1{Eq.~(\ref{eq:#1})}

\def\Figref#1{Fig.~\ref{fig:#1}}

\def\al{\alpha}
\def\be{\beta}
 
 \def\De{\Delta}

\def\la{\lambda}

\def\lag{{\mathcal L}}

\def\mathscr{\mathcal}

\def\3d{3-D}

\def\ol#1{\overline{#1}}
\def\amm{a.m.m.}

\def\s#1{\setbox0=\hbox{$#1$}  
   \dimen0=\wd0     
   \setbox1=\hbox{/} \dimen1=\wd1  
   \ifdim\dimen0>\dimen1   
      \rlap{\hbox to \dimen0{\hfil/\hfil}} 
      #1     
   \else     
      \rlap{\hbox to \dimen1{\hfil$#1$\hfil}} 
      /      
   \fi}      %

\begin{document}

\title{Electromagnetic moments of quasi-stable baryons}



\author{T. Ledwig}
\affiliation{Institut f\"ur Kernphysik, Universit\"at Mainz, D-55099 Mainz, Germany}

\author{J. Martin-Camalich}
\affiliation{Department of Physics and Astronomy, University of Sussex, BN1 9Qh,
    Brighton, UK.\\
Departamento de Fisica Teorica and IFIC, Universidad de Valencia-CSIC, Spain}

\author{V. Pascalutsa}
\affiliation{Institut f\"ur Kernphysik, Universit\"at Mainz, D-55099 Mainz, Germany}

\author{M. Vanderhaeghen}
\affiliation{Institut f\"ur Kernphysik, Universit\"at Mainz, D-55099 Mainz, Germany}

\begin{abstract}
We address electromagnetic properties of quasi-stable baryons in the context
of chiral extrapolations of lattice QCD results.
For particles near their decay threshold we show that the application of a small external magnetic field
changes the particle's energy in a non-analytic way. Conventional electromagnetic moments are only well-defined when the background field $B$
satisfies $|eB|/2M_*\ll|M_*-M-m|$ where $M_*$ is the mass of the resonance and
$M$, $m$ the masses of the decay products.
An application of this situation is the chiral extrapolation of $\Delta(1232)$-isobar electromagnetic properties. We discuss such an
extrapolation of the $\Delta(1232)$-isobar magnetic dipole, electric quadrupole and magnetic octupole
moments by a covariant chiral effective field theory.
\end{abstract}

\maketitle


\section{Introduction}
Conventionally, the energy shift of a spin 1/2 particle in an external magnetic field $\vec{B}$ is
given by:
\beq
\Delta E = -\vec{\mu}\cdot \vec{B}\,\,\,,
\eqlab{DeltaEnergy}
\eeq
which defines the particle's static magnetic moment $\vec{\mu}$. However, for
resonances near their decay threshold it was
shown that the
application of a small external magnetic field changes the resonance's
energy in a non-analytic way \cite{QSparticles}. As a consequence,
electromagnetic moments of a resonance
with mass $M_*$ and decay product masses $M$ and $m$ are only well defined if the condition 
\beq
\eqlab{cond1}
|eB|/2M_*\ll|M_*-M-m|
\eeq
is met. In order to use the approximation \Eqref{DeltaEnergy} for resonances
the condition expressed by \Eqref{cond1} has to be fulfilled. We sketch the derivation of this relation in the second section of
this report. 

Fields in which one encounters the above situation are lattice QCD
calculations and field theories.
Lattice QCD begin to obtain results for the $\Delta(1232)$-isobar
electromagnetic moments
\cite{Aubin(2008):LatticeDelta,Alexandrou(2009):LatticeDelta} where the
lightest pion mass is around $m_\pi\simeq 300$ MeV and a chiral
extrapolation to the physical point is needed. In doing so, one crosses the threshold
value $m_\pi=M_\Delta-M_N$, with $M_\Delta-M_N$ the $\Delta(1232)$-nucleon
mass gap, for which the right hand side of \Eqref{cond1} is zero.
In the third section we discuss the chiral behavior of
$\Delta(1232)$-isobar electromagnetic moments by a manifestly covariant chiral
effective field theory \cite{DeltaCHIRALbehavior}. Explicitly studied are the $\Delta(1232)$-isobar magnetic dipole, electric quadrupole and magnetic octupole moments where cusps
and singularities reflect the non-applicability of \Eqref{cond1} at the point $m_\pi=M_\Delta-M_N$.

\section{Anomalous magnetic moment and energy shift of a resonance}
We consider a model of two spin 1/2 fields $\Psi$ and $\psi$ of masses
$M_*$ and $M$, respectively, interacting with a scalar field $\phi$ of mass $m$ by:
\beq
\lag_{\mathrm{int}} = g\, \Big(\, \ol{\Psi} \,\psi\,\phi  + \ol \psi \,{ \Psi}\, \phi^\ast \Big) ,
\eeq 
where $g$ is a Yukawa coupling constant. The leading order electromagnetic vertex
corrections to the anomalous magnetic moment (\amm) of
$\Psi$ for charged $\phi$ and $\psi$ are Feynman-diagrams of the
type (D3) and (D4) that are depicted in the next section. For the above Lagrangian the double
lines in these graphs denote the $\Psi$, the single lines the $\psi$ and the dotted lines the $\phi$.
In the following, we concentrate on graph (D3) a
discussion of graph (D4) is analogous.

Coupling the photon minimally to $\phi$ yields the following \amm~$\kappa_*$ together with the conventional energy shift $\Delta \tilde{E}$:
\beq
\eqlab{kappa1}
\kappa_\ast =  \frac{2 g^2}{(4\pi)^2}\int_0^1 \! dx  \,\frac{- (r+x)\,x (1-x) }{x \mu^2 - x (1-x) + (1-x) r^2}\;\;\;\;\;,\;\;\;\;\;\Delta \tilde{E} = -\frac{\kappa_*}{2} \tilde{B}\,\,\,,
\eeq
with $r=M/M_\ast$, $ \mu=m/M_\ast $ and the dimensionless variables: $\tilde{B}=\frac{e B_z}{M_\ast^2}$  and $\De \tilde{E}=\frac{\De E}{M_\ast} + \frac{1}{2} \tilde B $.
In the left panel of \Figref{kappaANDdeltaE} we plot $\kappa_*$ as a function of the mass parameter $\mu$. We obtain a singularity at the point $m=M_*-M$ ($\mu=1-r$),
i.e. where $m$ is equal to the mass gap of $\Psi$ and $\psi$. The same singularity occurs for the graph (D4) with
different overall factors. Since an
infinite \amm~would imply an infinite energy shift of the resonance in an
external magnetic field, this singularity can not be physical. 
\begin{figure}{}
\includegraphics[height=.2\textheight]{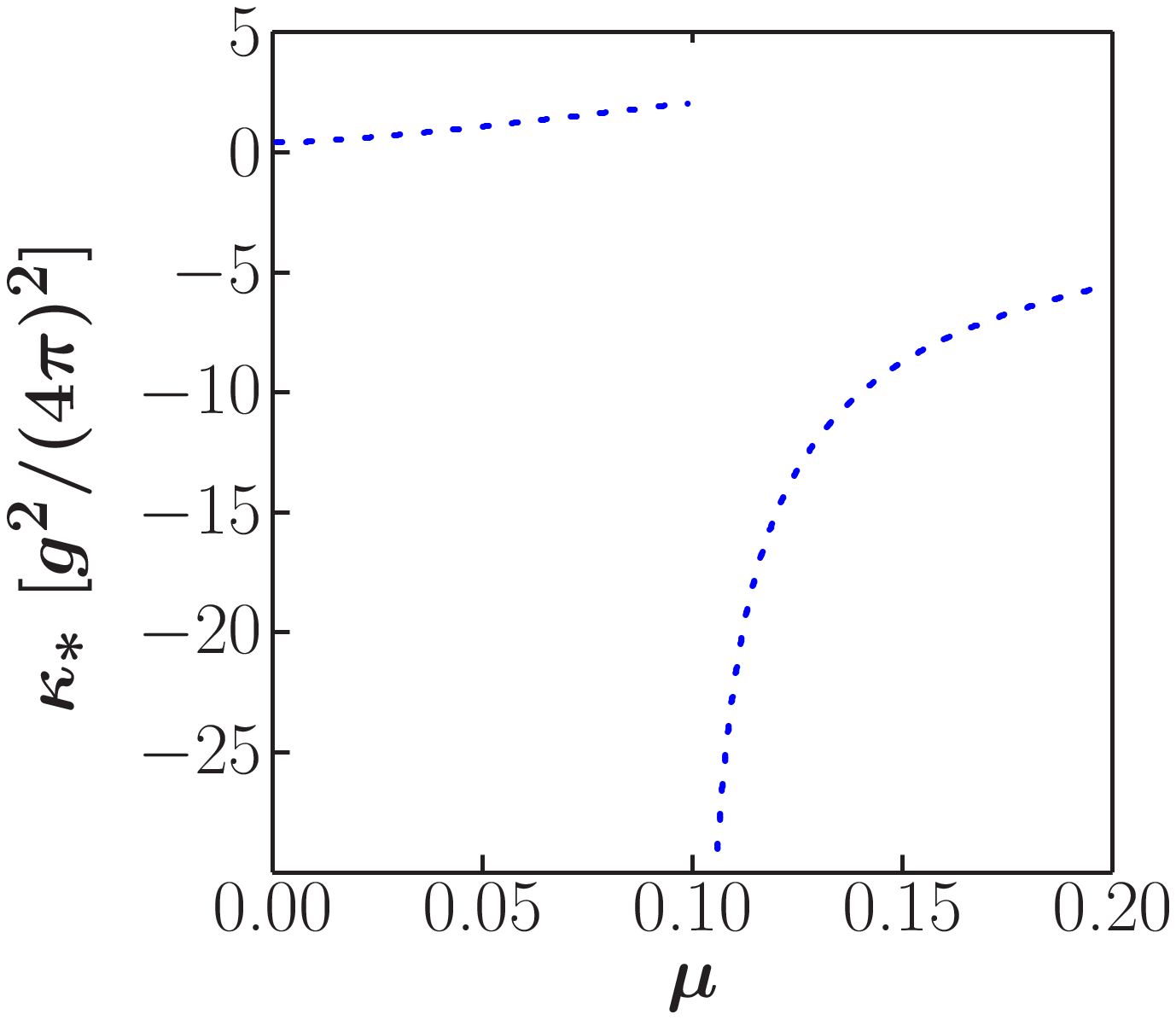}\,\,\,\includegraphics[height=.2\textheight]{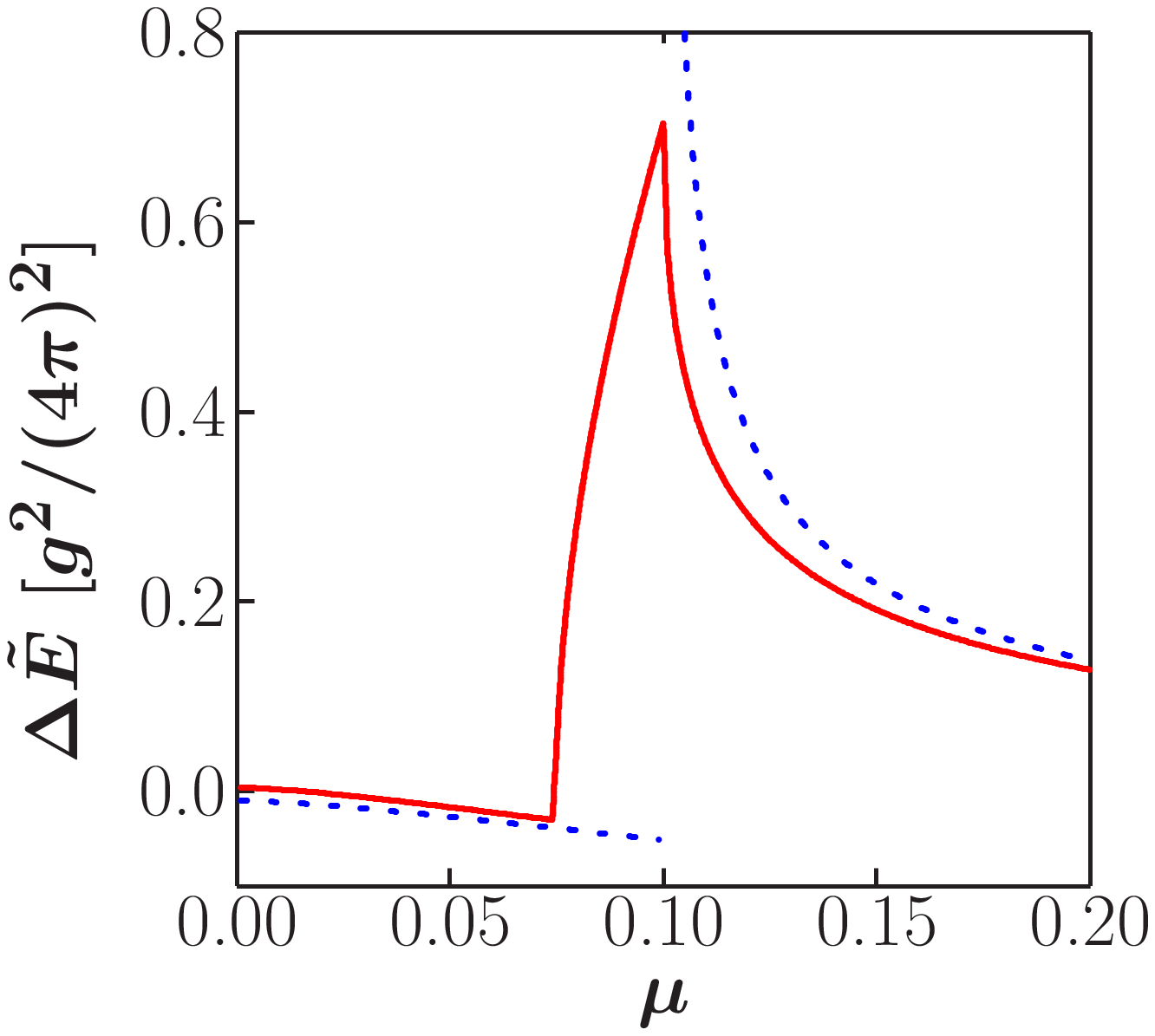}\,\,\,\includegraphics[height=.2\textheight]{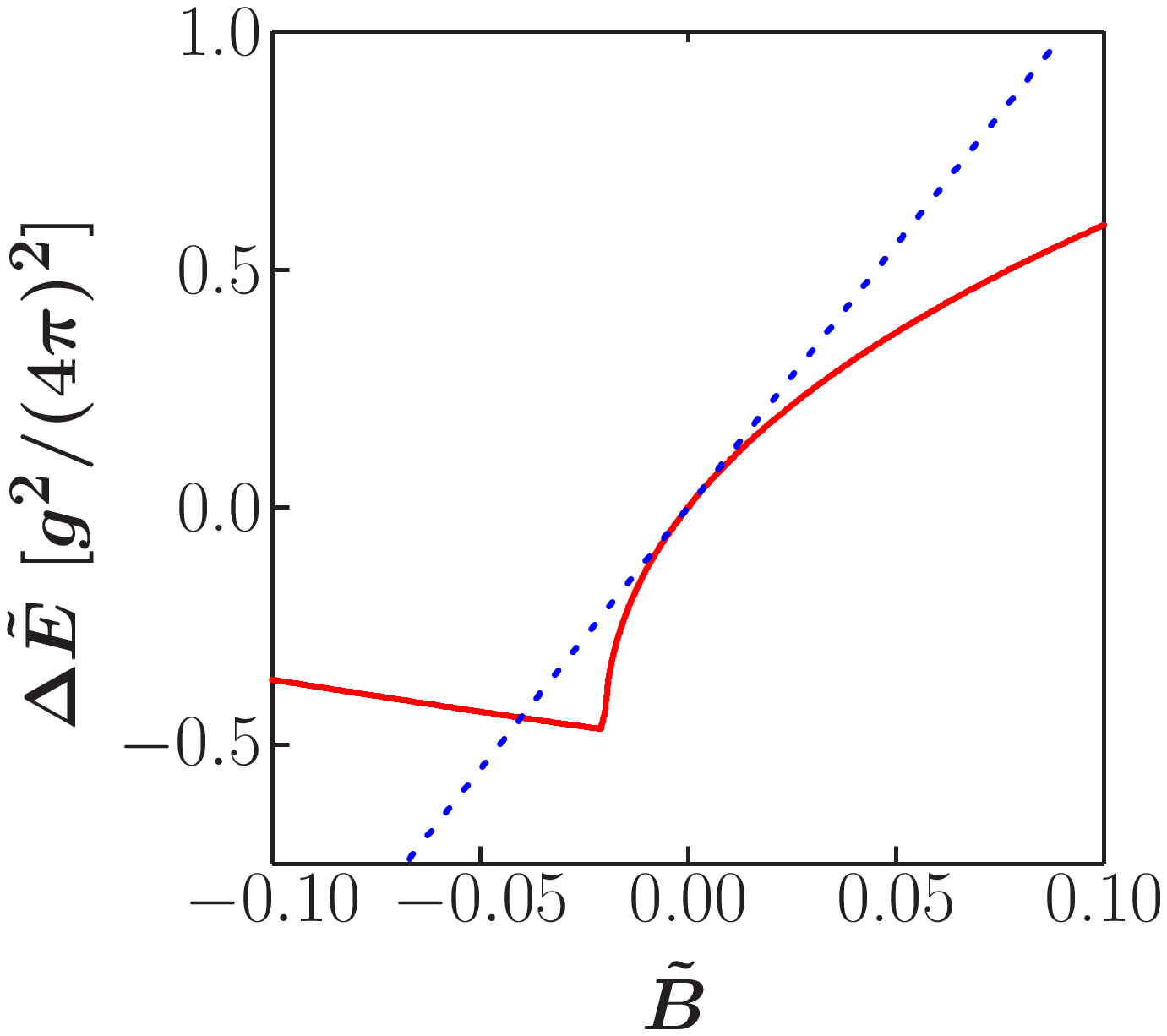}
\caption{Left panel: The \amm~contribution $\kappa_*$ as function of
  $\mu=m/M_*$ for fixed value $r=M/M_*=0.9$. The dashed blue curve depicts the real
  part. The imaginary part exhibits a similar singularity that is not shown. Middle panel: Energy
  shift of the particle $\Psi$ as a function of $\mu$ in a constant field of
  $\tilde{B}=0.05$ and $r=0.9$. The dashed blue curve depicts the linear
  approximation \Eqref{kappa1} while the solid red line depicts the solution of
  \Eqref{shift1}. Right panel: Same quantities as in the middle panel this
  time plotted for constant $\mu=0.09$ and variable $\tilde{B}$.}
\figlab{kappaANDdeltaE}
\end{figure}
The nature of this singularity can be seen by calculating the energy shift of the
resonance $\Psi$ with the background field technique of
\cite{Sommerfield:electronAMM}.

We consider a constant magnetic field in z-direction, $\vec B=B \vec{e}_z$, 
together with a $\Psi$ spin projection of $s_z=+1/2$. The resulting
energy-shift for the case $\Psi$ and $\phi$ are
positively charged is \cite{QSparticles}:
\beq
\De \tilde{E} =  \frac{g^2}{(4\pi)^2}\Big\{ (r+\alpha)\, 
(\Omega+\mathcal{A})-[(r+\alpha)\, (\Omega +\mathcal{A})
]_{\tilde{B}=0}\Big\}\;\;\;,
\eqlab{shift1}
\eeq
where the $\mathcal{A}$ contribution is analytic in $\tilde{B}$ and the $\Omega$ is non-analytic :
\beq
\mathcal{A} =-2+\alpha\ln r^2+\be\ln\mu^2
-\,\frac{\mu^2(1-\ln\mu^2)-r^2(1-\ln r^2)}{2(\al+r) (1-\tilde{B})}\;\;\;\;\;,\;\;\;\;\; \Omega=\lambda\, \ln\frac{( \alpha +\lambda)(\be+\la) }{(\alpha -\lambda)(\be-\la)}.
\eeq
with 
\beq
\alpha  =  \frac{1}{2(1-\tilde{B})}\left(1+r^2-\mu^2-\tilde{B}\right)\;\;\;,\;\;\;
\be  =  \frac{1}{2(1-\tilde{B})}\left(1-r^2+\mu^2-\tilde{B}\right)\;\;\;,\;\;\;
\lambda  =  \Big[ \alpha^{2}-r^{2}/(1-\tilde{B}) \Big]^{1/2}\;\;\;.
\eeq
In the middle panel of \Figref{kappaANDdeltaE} we plot \Eqref{kappa1} and \Eqref{shift1}. We see that the linear approximation
\Eqref{kappa1} does not accurately reproduce the energy shift of
$\Psi$ in the vicinity of the threshold $m=M_*-M$. Furthermore, plotting \Eqref{shift1} as a function of $\tilde{B}$,
right panel of \Figref{kappaANDdeltaE}, reveals a cusp that occurs at
$\tilde{B}=0$ for the case $m=M_*-M$. Hence, a definition of the magnetic
moment as the derivative at $\tilde{B}=0$ is not possible.
From the $\lambda$ term in $\Omega$ we can see that the energy shift $\Delta
\tilde{E}$ can only be expanded, i.e. a magnetic moment defined by \Eqref{DeltaEnergy}, when the condition \Eqref{cond1} is fulfilled. 
The energy shift $\Delta\tilde{E}$ on the other hand is the physical
observable, accessible e.g. in lattice QCD calculations.

\section{Chiral behavior of $\Delta(1232)$ electromagnetic properties}

For the $\Delta(1232)$-isobar, a Lorentz-covariant decomposition
of the electromagnetic $\gamma\Delta\Delta$ vertex with explicit
electromagnetic gauge invariance involves four form factors
\cite{Nozawa(1990):DeltaMATELEM}:
\begin{equation}
\langle\Delta^{\prime}|V^{\mu}|\Delta\rangle  = -\bar{u}_{\alpha}(p^{\prime})\Big\{\left[F_{1}^{*}(Q^{2})\gamma^{\mu}+\frac{i\sigma^{\mu\nu}q_{\nu}}{2M_{\Delta}}F_{2}^{*}(Q^{2})\right]g^{\alpha\beta}\nonumber+\left[F_{3}^{*}(Q^{2})\gamma^{\mu}+\frac{i\sigma^{\mu\nu}q_{\nu}}{2M_{\Delta}}F_{4}^{*}(Q^{2})\right]\frac{q^{\alpha}q^{\beta}}{4M_{\Delta}^{2}}\Big\}u_{\beta}(p)\,\,\,,\eqlab{onephoton}
\end{equation}
where $u_{\alpha}(p)$ is the Rarita-Schwinger spinor for a spin-3/2
state with mass $M_\Delta$. 
At $Q^{2}=0$, these form factors are related to the $\Delta(1232)$-isobar magnetic dipole $\mu_{\Delta}$, electric quadrupole $\mathcal{Q}_{\Delta}$
and magnetic octupole $\mathcal{O}_{\Delta}$ moments: 
\begin{equation*}
\mu_{\Delta}  = \frac{e}{2M_{\Delta}}\left[e_{\Delta}+F_{2}^{*}(0)\right]\,\,\,\,\,,\,\,\,\,\,\mathcal{Q}_{\Delta}
=
\frac{e}{M_{\Delta}^{2}}\left[e_{\Delta}-\frac{1}{2}F_{3}^{*}(0)\right]\,\,\,\,\,,\,\,\,\,\,
 \mathcal{O}_{\Delta}  =  \frac{e}{2M_{\Delta}^{3}}\left[e_{\Delta}+F_{2}^{*}(0)-\frac{1}{2}\left(F_{3}^{*}(0)+F_{4}^{*}(0)\right)\right]\,\,\,\,.
\end{equation*}
To study the chiral behavior of these moments, we use the chiral effective Lagrangian given by the B$\chi$PT Lagrangian of \cite{Gasser(1988):ChPT}
where the $\Delta(1232)$-isobar is included with the $\delta$-power
counting scheme of \cite{Pascalutsa(2003):deltaPOWERCOUNTING}. The explicit
Lagrangian consisting of pion, nucleon, $\Delta(1232)$-isobar and photon
fields can be found in
\cite{Pascalutsa(2005):DeltaMDM,DeltaLagrangian}. The
power counting breaking terms as found in \cite{Gasser(1988):ChPT} are
treated by the renormalization precription of \cite{Gegelia(1999):EOMS},
i.e. in addition to the divergent loop contributions also the finite power
counting breaking terms are subtracted.

In \Figref{fig:DiagramsDELTA} we show the
Feynman diagrams that contribute to the $\Delta(1232)$-isobar electromagnetic
moments at the order $p^3$ and $p^4/\Delta$ with $\Delta = M_\Delta - M_N$. 

\begin{figure}[h]
\includegraphics[scale=0.25]{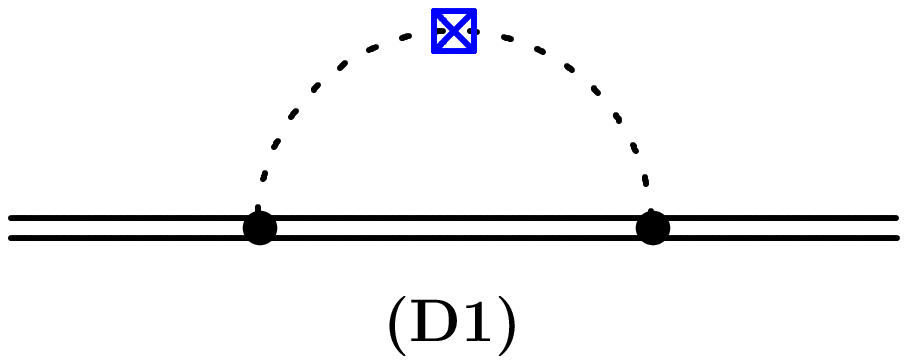}\;\;\includegraphics[scale=0.25]{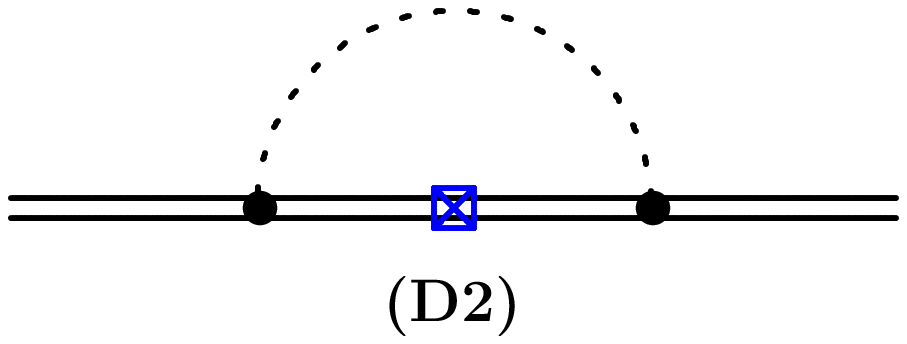}\;\;\includegraphics[scale=0.25]{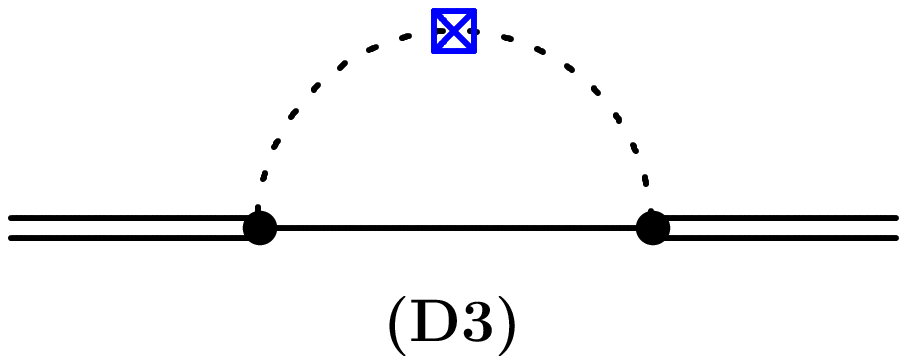}\;\;\includegraphics[scale=0.25]{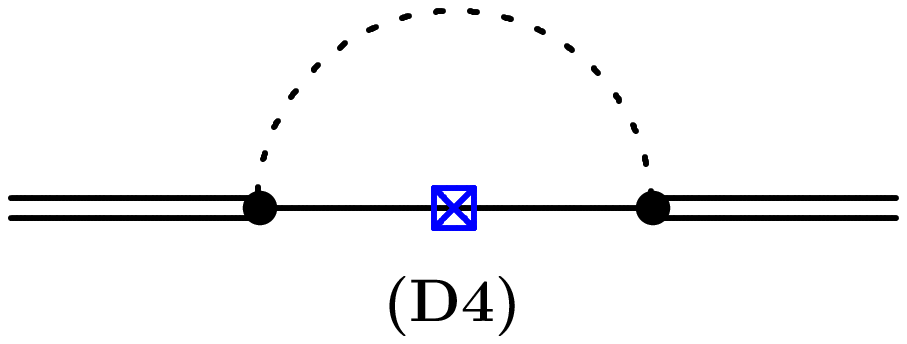} 
\end{figure}
\begin{figure}[h]
\includegraphics[scale=0.25]{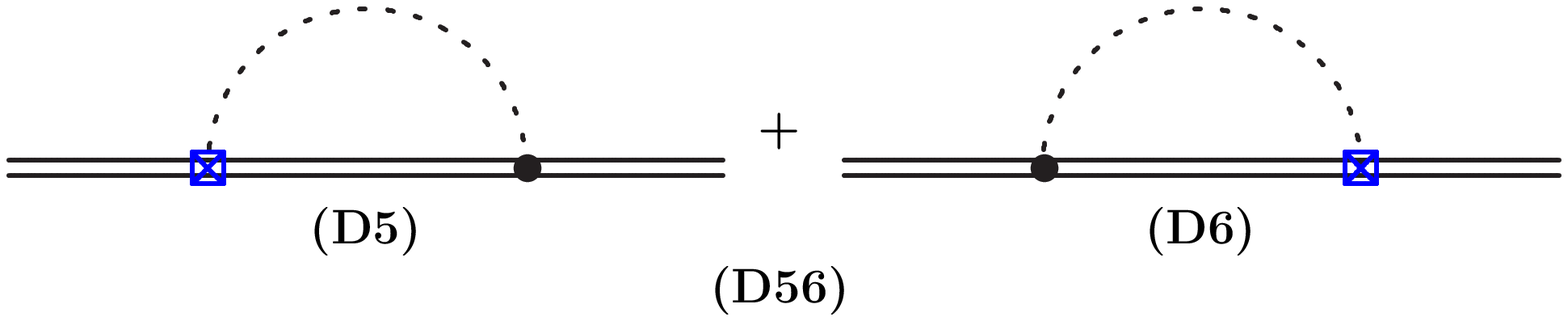}\,\,\includegraphics[scale=0.25]{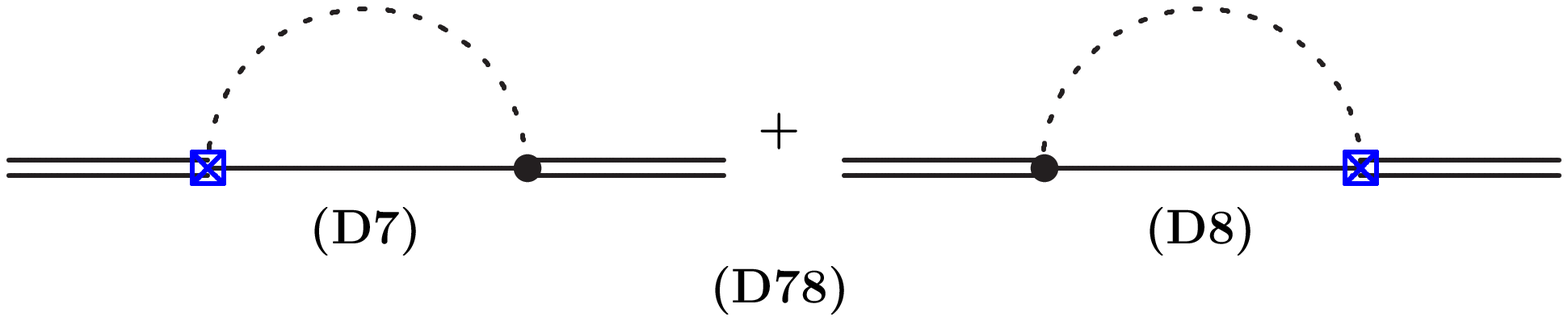}\\
\caption{Diagrams contributing to $\Delta(1232)$
electromagnetic moments.
Solid-single lines represent nucleons, solid-double lines the $\Delta(1232)$-isobar
and dashed lines the pions. The photon coupling is denoted by blue squares while the $N\pi$ and $\Delta\pi$ vertices by black dots. }
\figlab{fig:DiagramsDELTA}
\end{figure}

In \Figref{DeltaMoments} we show results of our investigation
\cite{DeltaCHIRALbehavior}. Depicted are the $\Delta(1232)^+$ magnetic dipole,
electric quadrupole and magnetic octupole moments as functions of the pion
mass squared $m_\pi^2$. The lattice
QCD data points are taken from \cite{Aubin(2008):LatticeDelta} green
triangles and \cite{Alexandrou(2009):LatticeDelta} orange rectangles. The electromagnetic moments exhibit cusps and
singularities at the threshold value $m_\pi=M_\Delta -M_N$, with $M_N$ as the
nucleon mass, that stem from the graphs (D3) and (D4). The cusp in the magnetic moment was already seen in
Ref. \cite{Pascalutsa(2005):DeltaMDM}. As discussed in the previous section,
this behavior is a consequence of the non-fulfillment of
condition \Eqref{cond1} for that $m_\pi$ value. At this point one can not
find any weak magnetic field which enables to define properly the
electromagnetic properties of the $\Delta(1232)$-isobar in the conventional way.

For lattice QCD calculations of electromagnetic moments there exist the three point
function and the background field methods. In the case of the
background field technique, the periodicity condition gives a lower bound for the applied magnetic field
strength \cite{Aubin(2008):LatticeDelta,Lee(2005):MagStrength}. Hence, it
exists a $m_\pi$ region around the threshold $m_\pi=M_\Delta-M_N$ in which \Eqref{cond1} is not met.
\begin{figure}[h]
\includegraphics[scale=0.45]{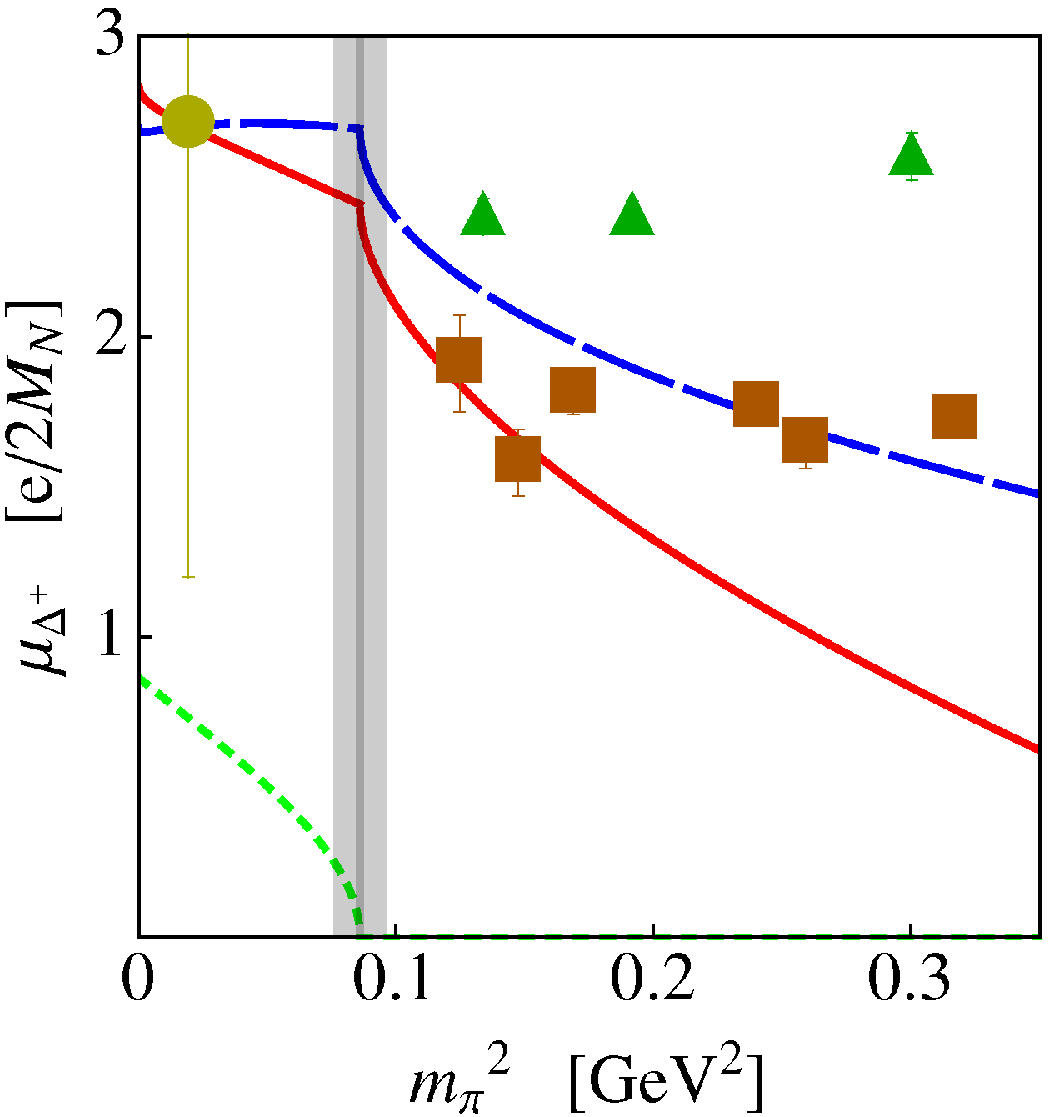}\,\,\,\,\,\includegraphics[scale=0.45]{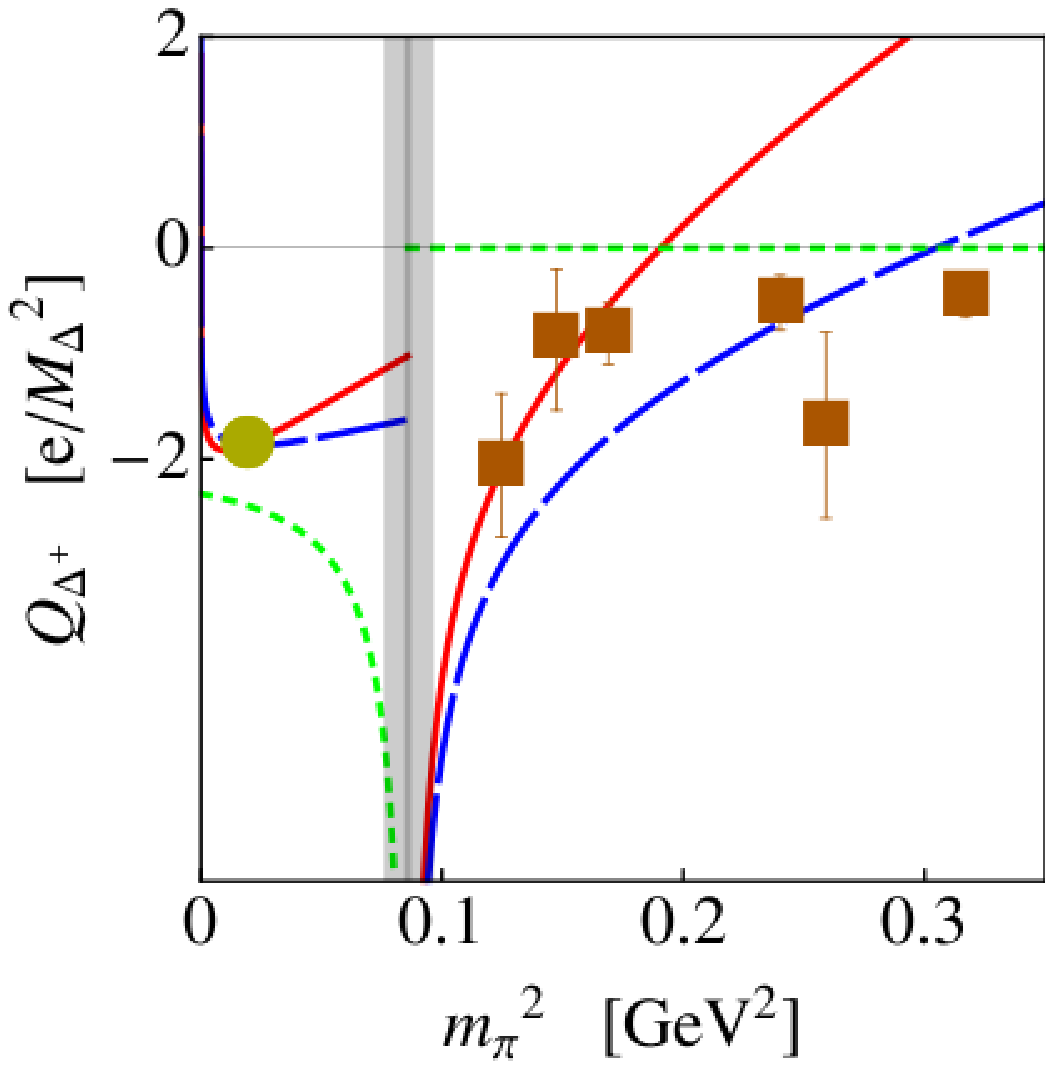}\,\,\,\,\,\includegraphics[scale=0.45]{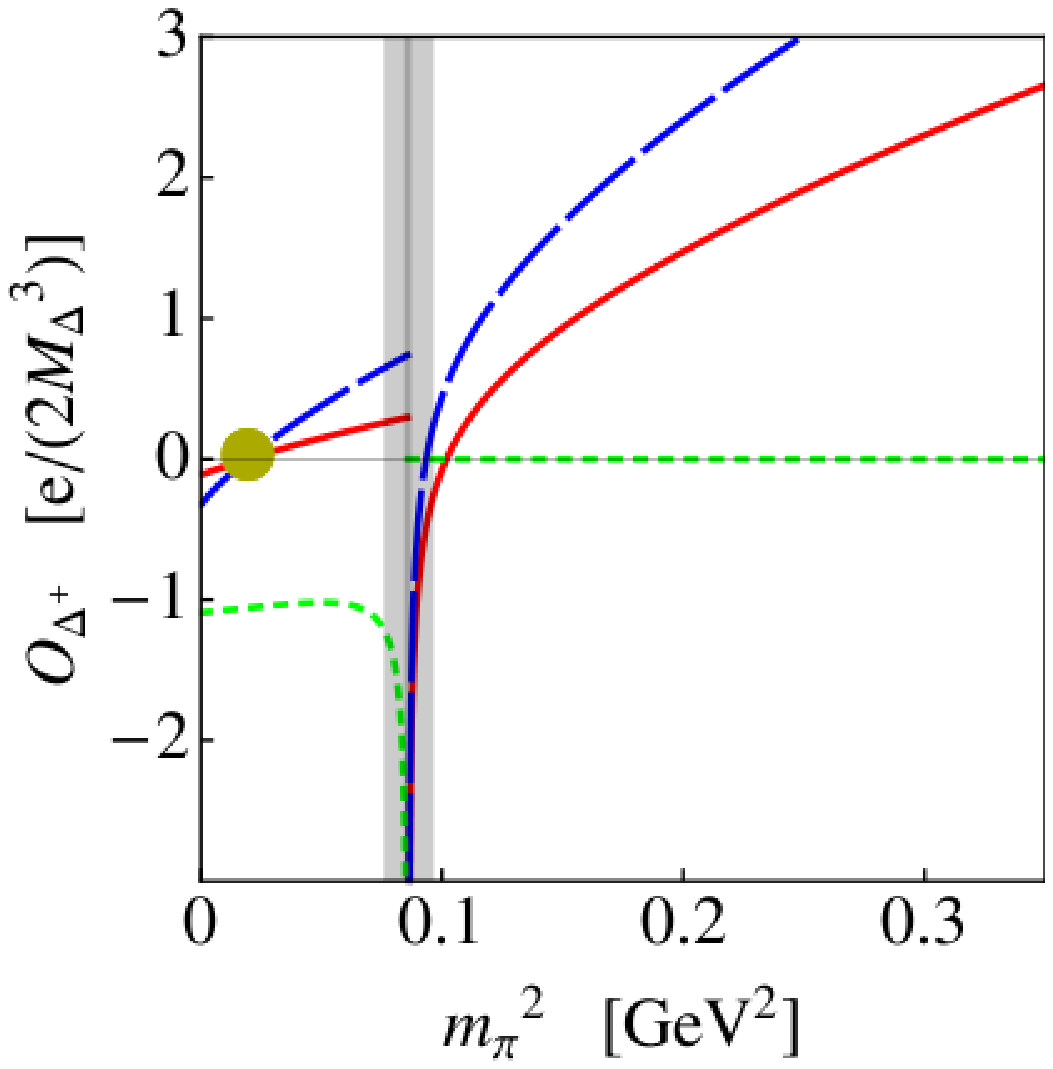}
\caption{The magnetic dipole $\mu_{\Delta^{+}}$, electric quadrupole
  $\mathcal{Q}_{\Delta^{+}}$ and magnetic octupole $\mathcal{O}_{\Delta^{+}}$
  moments of the $\Delta^{+}(1232)$. The curves correspond to: red solid and
  blue long-dashed curves show our results with and without inclusion of $\gamma\Delta\Delta$
non-minimal couplings, respectively, and are constraint to values at the physical pion
  mass; green short-dashed curve to the
imaginary parts. The yellow circle corresponds to the used values at the
physical point: the experimental value
$\mu_{\Delta^{+}}=\left(2.7\pm1.5\right)\,\mu_{N}$, a large-$N_c$ estimate
$\mathcal{Q}_{\Delta^+}=-1.87\,\,e/M_\Delta^2$ and $\mathcal{O}_{\Delta^+}=0$.
The lQCD data of \cite{Aubin(2008):LatticeDelta} are denoted by
green triangles while those of \cite{Alexandrou(2009):LatticeDelta}
are depicted by orange rectangles. The grey bands are described in
the text.}
\figlab{DeltaMoments}
\end{figure}
To give explicit situations, we take two spatial lattices of $L=24$
with spacing $a^{-1}=2$ GeV and $L=32$ with spacing $a^{-1}=1$ GeV and
implement the magnetic field by $eBa^{2}=2\pi/L^{2}$ \cite{Aubin(2008):LatticeDelta}.
Further, we take \Eqref{cond1} to be unity, i.e. a completely
non-fulfillment of this relation, and solve for that region. There higher
order $\vec{B}$-terms in the energy $E(\vec{B})$ cannot
be neglected and a static electromagnetic moment is not well defined in the
traditional way. We see that for the given lattices this region ranges
from $m_{\pi}=275.3$ MeV to $m_{\pi}=310.7$ MeV for the coarse lattice and
from $m_{\pi}=290.5\sim295.5$ MeV for the 
finer lattice. We represent these two regions as grey bands in
\Figref{DeltaMoments}.

A similar problem could also affect lattice calculations of resonance \amm~by the
three point function method. Here the results are limited by finite $Q^2$ and
an extrapolation to $Q^2=0$ needs to be done. Qualitatively, a finite $Q^2$ would enter as an
additional energy paramter and the singularities would be shifted away from
the point $Q^2=0$ for pion masses other than $M_\Delta-M_N$. Hence, for
a given $m_\pi$ one would get a finite $Q^2$ value for which the form factor is singular. For practical lattice calculations
this could mean that one extrapolates across this singularity to $Q^2=0$ when
all data points are on the right of the singularity.

\section{Conclusion}
We addressed electromagnetic properties of quasi-stable particles. For such
particles we showed that the application of a small external
magnetic field changes the particle's energy in a non-analytic
way. Static electromagnetic moments are only well defined when the
condition \Eqref{cond1} is fulfilled \cite{QSparticles}. Explicit situations where
\Eqref{cond1} could be violated are lattice QCD calculations of
electromagnetic properties and their chiral extrapolations by means of
effective-field theories. We discussed such an extrapolation for the $\Delta(1232)$-isobar
electromagnetic moments by means of a covariant chiral effective field theory
\cite{DeltaCHIRALbehavior}. The non-fulfillment of \Eqref{cond1} is reflected
by cusps and singularities at the point $m_\pi=M_\Delta-M_N$. Modern lattice QCD
results are about to approach the pion mass region where this condition 
applies and the presented techniques can be used.

\section*{Acknowledgments}
The work of TL was partially supported
by the Research Centre \char`\"{}Elementarkraefte und Mathematische
Grundlagen\char`\"{} at the Johannes Gutenberg University Mainz. JMC acknowledges the MEC contract FIS2006-03438, the EU Integrated Infrastructure Initiative Hadron Physics Project contract RII3-CT-2004-506078 and the Science and Technology Facilities Council [grant number ST/H004661/1] for support.

\end{document}